\documentclass[prc,twocolumn,showpacs,preprintnumbers,amsmath,amssymb,superscriptaddress,floatfix]{revtex4}
\usepackage{amsfonts}%
\usepackage{graphicx}%
\newcommand{\half}{\mbox{${\textstyle \frac{1}{2}}$}}

\begin{document}
\title{On the Coupling Constant for $\mathbf{N^*(1535)N\rho}$}
\author{Ju-Jun Xie} \email{xiejujun@mail.ihep.ac.cn}
\affiliation{Institute of High Energy Physics and Theoretical
Physics Center for Science Facilities, CAS, Beijing 100049, China}
\affiliation{Graduate University of Chinese Academy of Sciences,
Beijing 100049, China}
\author{Colin Wilkin} \email{cw@hep.ucl.ac.uk}
\affiliation{Physics and Astronomy Department, UCL, London WC1E 6BT,
UK}

\author{Bing-Song Zou} \email{zoubs@mail.ihep.ac.cn}
\affiliation{Institute of High Energy Physics and Theoretical
Physics Center for Science Facilities, CAS, Beijing 100049, China}
\affiliation{Center of Theoretical Nuclear Physics, National
Laboratory of Heavy Ion Accelerator, Lanzhou 730000, China}

\begin{abstract}
The value of the $N^*(1535)N\rho$ coupling constant $g_{N^*N\rho}$
derived from the $N^*(1535) \to N\rho \to N \pi \pi$ decay is
compared with that deduced from the radiative decay $N^*(1535) \to N
\gamma$ using the vector-meson-dominance model. On the basis of an
effective Lagrangian approach, we show that the values of
$g_{N^*N\rho}$ extracted from the available experimental data on the
two decays are consistent, though the error bars are rather large.
\end{abstract}


\maketitle
%
%
\section{INTRODUCTION}
\label{Introduction}

The experimental database on the production of the $\eta$ meson in
nucleon-nucleon scattering near threshold has expanded significantly
in recent years. In addition to measurements of $pp\to pp\eta$ total
cross sections and angular distributions~\cite{eta_exp}, there are
analyzing powers~\cite{Rafal} and full Dalitz plots~\cite{Pauly}.
Total cross sections are also available for the $pn\to d\eta$ and
$pn\to pn\eta$ reactions~\cite{Calen}.

In response to this wealth of data there have been a large number
of theoretical investigations of $\eta$ production in both
proton-proton and proton-neutron collisions. Most of these have
been within the framework of meson-exchange models, where a
$N^*(1535)$ resonance or other nucleon isobar is excited through
the exchange of a single meson, with the $\eta$-meson being formed
through the decay of the isobar. There are differences in the
literature on how to treat the initial and final state
interactions but the major controversies are connected with which
meson exchanges are deemed to be important.

The large ratio of the production of the $\eta$ in proton-neutron
compared to proton-proton collisions suggests that isovector
exchange plays the major role. However, some
authors~\cite{bati,naka,shyam07} find pseudoscalar ($\pi$ and
$\eta$) exchanges to dominate, with no significant contribution from
the $\rho$. In contrast, others~\cite{geda,san,faldt,naka02,vetter}
claim that $\rho$-meson exchange plays an important and possibly
dominant role. This disagreement is generated principally by the
uncertainty in the size of the $N^*(1535)N\rho$ coupling and it is
the purpose of this present note to compare the values of the
coupling constant derived from the $N^*(1535)\to N\pi\pi$ and
$N^*\to N\gamma$ decays.

The situation is further complicated by the variety of forms
chosen for the $N^*(1535)N\rho$ coupling in these different works.
In the vector meson dominance model (VMD), it is assumed that this
coupling is proportional to that for the electromagnetic
$N^*(1535)N\gamma$. It should be noted that in this approach the
tensor $\sigma_{\mu \nu}$ coupling automatically satisfies the
associated gauge invariance constraint~\cite{naka}. In contrast,
the vector $\gamma_5 \gamma_{\mu}$ coupling violates gauge
invariance when the $\rho$-meson is replaced by a
photon~\cite{naka,naka02}. As an alternative, Riska and
Brown~\cite{riska} suggested a vertex of the form
$\gamma_5[\gamma^{\mu}p_{\rho}^2-(M_{N^*}+m_N)p^{\mu}_{\rho} ]$,
where $p_{\rho}$ is the four-momentum of the $\rho$ meson. This
avoids the gauge invariance problem while keeping the
$\gamma_{\mu}$ term, but this coupling vanishes when used in
connection with the VMD approach. In principle both vector and
tensor couplings are needed and their relative importance has to
be decided by experiment.

Working within an effective Lagrangian approach, we have
investigated the influence of the $N^*(1535)N \rho$ coupling
constant on both the $N^*(1535) \to N \rho^0 \to N\pi^+ \pi^-$ and
the $N^*(1535) \to N\rho^0 \to N\gamma$ decays. In
Sect.~\ref{Formalism}, we present the formalism and ingredients
necessary for our estimations. Although in one case the
$\rho$-meson is essentially real while in the other it has zero
mass, we show in Sect.~\ref{Results} that consistent values of the
coupling constant can be obtained from the available experimental
data on the two decay channels, though the uncertainties are still
quite large.
%
%
\section{FORMALISM AND APPLICATION}
\label{Formalism}

\begin{figure}
\includegraphics[scale=1.3]{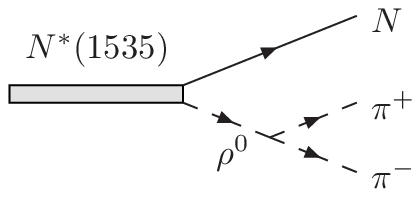} \\
\vspace{0.8cm}
\includegraphics[scale=1.3]{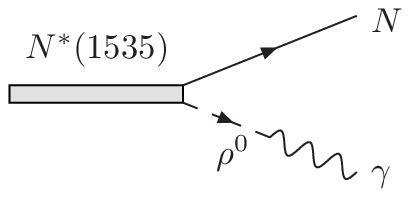}
\caption{Feynman diagrams considered for the $N^*(1535) \to N \rho^0
\to N \pi^+ \pi^-$ and $N^*(1535) \to N \rho^0 \to N \gamma$
decays.} \label{diagram}
\end{figure}

The basic Feynman diagrams for the two cascade decay modes
considered here are depicted in Fig.~\ref{diagram}. A Lorentz
covariant orbital-spin ($L$--$S$) scheme for $N^*NM$ couplings has
been developed in detail in Ref.~\cite{zouprc03} and, within that
scheme, one can easily derive the form of the effective $N^*(1535) N
\rho$ coupling. Since the $\rho$ is a vector meson, both $S$- and
$D$-wave couplings are possible but experiment shows that the
$D$-wave plays only an insignificant role in the $N^*(1535)\to
N\rho$ partial decay width~\cite{vrana,pdg2006}. We therefore retain
only the $S$-wave term with a Lagrangian of the form
\begin{align}
\mathcal{L}_{\rho N N^*} = ig_{N^* N\rho}\bar{u}_N \gamma_5
\left(\gamma_{\mu}-\frac{q_{\mu} \not\!q}{q^2}\right)
\vec{\tau}\cdot\vec{\rho}^{\mu} u_{N^*} +h.c.\,,\label{nsrho}
\end{align}
where $u_N$ and $u_{N^*}$ are the nucleon and $N^*(1535)$ spinors
and $q$ is the isobar four-momentum. The $\rho$-meson field
$\vec{\rho}_{\mu}$ is also a vector in isospin space and
$\vec{\tau}$ is the isospin operator in the baryon sector. It is
seen that this form is a particular linear combination of vector
and tensor couplings.

The finite size of the hadrons is taken into account through a form
factor which is normalized to unity at $p^2_{\rho}=m^2_{\rho}$.
Since only the $S$-wave is involved, this is taken to be monopole
type
\begin{equation}
F(p^2_{\rho})=\frac{\Lambda^2}{\Lambda^2+|p^2_{\rho}-m^2_{\rho}|}\,,
\label{sff}
\end{equation}
with a cut-off parameter $\Lambda$. In the case of t-channel
exchange, $p^2_{\rho}<m^2_{\rho}$, it leads to the more familiar
form
\begin{equation}
F(p^2_{\rho})=\frac{\Lambda^2_t-m^2_\rho}{\Lambda^2_t-p^2_\rho}\,,
\label{sff1}
\end{equation}
with $\Lambda^2_t=\Lambda^2+m^2_\rho$.

For the $\rho \pi \pi$ and $\rho \gamma$ couplings, we use the
standard Lagrangians~\cite{lixueqian, bauer, zhaoqiang},
\begin{eqnarray}
\mathcal{L}_{\rho \pi \pi}  &=& g_{\rho \pi \pi} (\vec{\pi}\times
\partial^{\mu} \vec{\pi}) \cdot \vec{\rho}_{\mu},
\label{rhopipi}\\
\mathcal{L}_{\rho \gamma} &=& \frac{e m^2_{\rho}}{f_{\rho}}
\rho_{\mu}^{0}A^{\mu}. \label{rhog}
\end{eqnarray}
where $\vec{\pi}$ and $A^{\mu}$ are the pion and electromagnetic
fields, respectively. The direct photon-vector coupling in Feynman
diagram language is reflected in the factor $e
m^2_{\rho}/f_{\rho}$.

The value of the $\rho \pi \pi$ coupling constant $g_{\rho \pi
\pi}$ can be deduced from the partial decay width
\begin{equation}
\Gamma_{\rho^0 \to \pi^+ \pi^-} = \frac{g^2_{\rho \pi \pi}}{6 \pi}
\frac{(p^{\,\text{cm}}_{\pi})^3}{m^2_{\rho}},
\end{equation}
where $p^{\,\text{cm}}_{\pi}$ is the momentum of one of the pions in
the rest frame of the $\rho$-meson. The experimental data then yield
$g^2_{\rho\pi\pi}/4\pi=2.91$.

Many photoproduction reactions have been successfully related to
ones involving the production or decay of vector mesons within the
vector meson dominance model. As a consequence, there are several
ways to evaluate the $\rho \gamma$ coupling constant but they
differ little from those given in the original Sakurai
compilation~\cite{sakurai} and we take $f^2_{\rho}/4\pi$ = 2.7.

The amplitude for the strong decay $N^*(1535) \to N \rho^0 \to N
\pi^+ \pi^-$ has the form
\begin{eqnarray}
&&\mathcal{M}_{N^* \to N \rho^0 \to N \pi^+ \pi^-}  =
ig_{\rho \pi \pi} g_{N^* N \rho} F(p^2_{\rho}) \times \nonumber  \\
&& \hspace{2mm}\bar{u}_N \gamma_5 \left(\gamma^{\mu}-\frac{\not\!q
q^{\mu}}{q^2}\right) u_{N^*}\, G^{\rho}_{\mu \nu}(p_{\rho})
(p^{\nu}_2 - p^{\nu}_3)\,,
\end{eqnarray}
Here $G_{\mu \nu}^{\rho}(p_{\rho})$ is the $\rho$-meson
propagator,
\begin{equation}
G_{\mu \nu}^{\rho}(p_{\rho}) = -i \frac{g^{\mu \nu}-p_{\rho}^{\mu}
p_{\rho}^{\nu}/p^2_{\rho}}{p^2_{\rho} - m^2_{\rho} + i m_{\rho}
\Gamma_{\rho}} \,,
\end{equation}
where $\Gamma_{\rho}$ is the total $\rho$ decay width.

The partial decay width is related to the spin-averaged amplitude
through
\begin{eqnarray}
&&d\Gamma_{N^* \to N \rho^0 \to N \pi^+ \pi^-}  =
\overline{|\mathcal{M}_{N^* \to N \rho^0 \to N \pi^+ \pi^-}|^2}\times \nonumber \\
&&\hspace{3mm}\frac{m_N}{(2\pi)^5}\frac{d^3p_1d^3p_2d^3p_3}{4E_1E_2E_3}
\delta^4(M_{N^*}\!-\!p_1\!-\!p_2\!-\!p_3)\,, \label{pipidecay}
\end{eqnarray}
where $p_1$, $p_2$, $p_3$ and $E_1$, $E_2$, $E_3$ are the momenta
and energies of the nucleon, $\pi^+$, and $\pi^-$, respectively.

\begin{figure}
\includegraphics[scale=0.5]{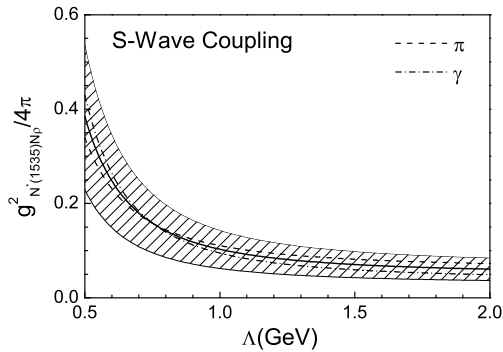}
\vspace{-2.5cm} \caption{Coupling constant $g^2_{N^*N\rho}/4\pi$
\emph{versus} the cut-off parameter $\Lambda$ for the pure $S$-wave
coupling case. The dashed curve was obtained from the $N^*\to
N\pi\pi$ decay whereas the dot-dashed one corresponds to the $N^*\to
N\gamma$ decay. The solid curve represents the average of the two
approaches with the shading showing the uncertainties arising from
the errors in the experimental input.}\label{ccs}
\end{figure}

The phase-space integration of Eq.~(\ref{pipidecay}) was evaluated
numerically and the values of the cut-off parameter of
Eq.~(\ref{sff}) and the $N^*(1535)N\rho$ coupling constant adjusted
to yield the experimental partial width of $(3.0\pm1.6)\,$MeV which
is obtained from the PDG values for the total decay width of $150\pm
25$ MeV and the branching ratio of $0.02\pm 0.01$~\cite{pdg2006}. In
Fig.~\ref{ccs} the value of $g_{N^*N\rho}^2/4\pi$ is shown as a
function of $\Lambda$ by the dashed curve. In view of the
uncertainty in the partial width, one should consider an error
corridor of $\pm 53\%$ around this curve.

Turning now to the radiative decay, the current best PDG estimates
of the helicity-\half\ decay amplitudes for the charged and neutral
$N^*(1535)$ are $A^{p \gamma}_{1/2}=0.090\pm
0.030\,(\textrm{GeV})^{-1/2}\,$ and $A^{n \gamma}_{1/2}=-0.046\pm
0.027\,(\textrm{GeV})^{-1/2}\,$, respectively~\cite{pdg2006}. These
lead to the corresponding isovector helicity-\half\ decay amplitude
of the $N^*(1535)$ as
\begin{equation}
A^{I=1}_{1/2} = \half\left(A^{p \gamma}_{1/2} - A^{n
\gamma}_{1/2}\right) = (0.068\pm0.020)\,(\textrm{GeV})^{-1/2}\,,
\label{arho}
\end{equation}
in terms of which the $N^*(1535) \to N \gamma$ partial decay width
for isovector photons becomes
\begin{equation}
\Gamma_{N^* N \gamma} = \frac{k^2}{\pi} \frac{m_N}{M_{N^*}}
\left(A_{1/2}^{I=1}\right)^{\!2}, \label{ngamma}
\end{equation}
where $k$ is the photon momentum in the $N^*$ rest frame.

The radiative decay width can be estimated within the VMD model by
applying the Feynman rules to Fig.~\ref{diagram}b. The resulting
matrix element is
\begin{eqnarray}
&&\nonumber\hspace{-4mm}\mathcal{M}_{N^*\to N \gamma} =
-i\frac{em^2_{\rho}}{f_{\rho}} g_{N^* N \rho} F(p^2_{\rho})
G^{\rho}_{\mu \nu}(p_{\rho})\,\varepsilon^{\nu}(k)\\
&&\hspace{1cm}\times\bar{u}_N \gamma_5
\left(\gamma^{\mu}-\frac{\not\!q q^{\mu}}{q^2}\right) u_{N^*},
\end{eqnarray}
where $\varepsilon^{\nu}(k)$ is the polarization vector of photon.
The resulting decay width is
\begin{equation}
\Gamma_{N^* \to N \gamma} = \frac{g^2_{N^* N \rho}}{4\pi}
\frac{\alpha}{f^2_{\rho}/4 \pi} \frac{3k(m_N +
E_N)}{M_{N^*}(1+\Gamma_{\rho}^2/m_{\rho}^2)(1+m_{\rho}^2/\Lambda^2)^2}
\,, \label{zou}
\end{equation}
where $\alpha$ is the fine-structure constant and $E_N$ the energy
of the final nucleon. The numerical value of $A^{I=1}_{1/2}$ from
Eq.~(\ref{arho}) leads to the dot-dashed curve of Fig.~\ref{ccs},
which shows $g_{N^*N\rho}^2/4\pi$ \emph{versus} $\Lambda$ as derived
from the radiative decay. An uncertainty corridor must also be
associated with this curve because of the large error in the
radiative amplitude shown in Eq.(\ref{arho}).

The values of $g^2_{N^*N\rho}/4\pi$ extracted from the two decays
are mutually compatible within the error bars for the whole range of
$\Lambda$ from 0.5 to 2.0 GeV. From these two independent
measurements we deduce the average value and the corresponding
uncertainty corridor, as shown by the solid curve and the shaded
area in Fig.~\ref{ccs}.

We can derive analogous constraints from these data on the other
two commonly used forms for the $N^*(1535)N\rho$ coupling, i.e.,
pure vector or pure tensor which have, respectively, the
corresponding effective Lagrangians
\begin{eqnarray}
\mathcal{L}^{\text V}_{\rho N N^*} & = & ig_{N^* N\rho}\bar{u}_N
\gamma_5 \gamma_{\mu} \vec{\tau}\cdot\vec{\rho}^{\mu} u_{N^*}
+h.c.\,,\label{nsrhot}
\\
\mathcal{L}^{\text T}_{\rho N N^*} & = & i\frac{g_{N^*
N\rho}}{2m_N}\bar{u}_N \gamma_5 \sigma_{\mu \nu} \partial^{\nu}
\vec{\tau}\cdot\vec{\rho}^{\mu} u_{N^*} +h.c.\,.\label{nsrhov}
\end{eqnarray}

Since these two kinds of coupling involve both $S$-wave and
$D$-wave, a dipole form factor is used for the $N^*N\rho$ vertex:
\begin{equation}
F(p^2_{\rho})=\left (
\frac{\Lambda^2}{\Lambda^2+|p^2_\rho-m^2_{\rho}|} \right )^2\,.
\label{dff}
\end{equation}

\begin{figure}
\includegraphics[scale=0.5]{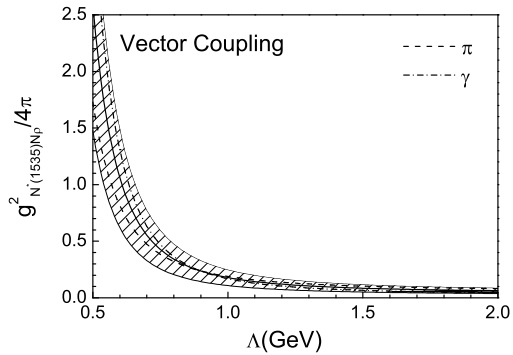}
\vspace{-2.5cm} \caption{As for Fig.~\ref{ccs} but for the pure
vector coupling case.} \label{ccv}
\end{figure}

\begin{figure}
\includegraphics[scale=0.5]{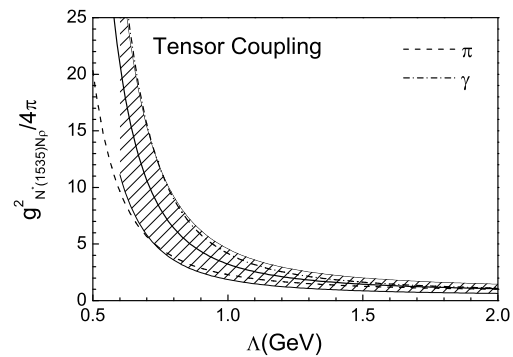}
\vspace{-2.5cm} \caption{As for Fig.~\ref{ccs} but for the pure
tensor coupling case.} \label{cct}
\end{figure}

The corresponding results are shown in Fig.~\ref{ccv} and
Fig.~\ref{cct}, respectively. For the pure vector coupling case, the
extracted values from the two decays are also agree within error
bars for the whole range of $\Lambda$ from 0.5 to 2.0 GeV, while for
pure tensor coupling this is only true for $\Lambda
> 0.6$ GeV.

%
%
\section{DISCUSSION AND CONCLUSIONS}
\label{Results}

In this short note we have compared the values obtained for the
$N^*(1535)N \rho$ coupling constant from experimental data on the
radiative and two-pion decays of $N^*(1535)$ resonance. For this
purpose we have used an effective Lagrangian approach combined with
the vector meson dominance model that links photoproduction
reactions to ones involving the $\rho$ and other vector mesons. With
a particular choice of the form of the $N^*(1535)N \rho$ vertex
($S$-wave coupling), we show in Fig.~\ref{ccs} that the two
determinations are quite compatible for a wide range of the cut-off
parameter $\Lambda$, especially if account is taken of the error
bands that arise from uncertainties in the input data. Typically one
would expect $\Lambda_t$ = $(\Lambda^2 + m^2_{\rho})^{1/2}$ to be of
the order of 1\,GeV/$c^2$~\cite{Bonn}, which falls well within the
domain of compatibility.

It is seen from Figs.~\ref{ccv} and \ref{cct} that the pure vector
and tensor forms of the coupling can also reproduce simultaneously
the data within the rather large error bars, though marginally worse
than the pure $S$-wave coupling of Fig.~\ref{ccs}. Both vector and
tensor forms are linear combinations of $S$-wave and $D$-wave
couplings. However, since both give only a small $D$-wave
contribution to $N^*(1535)\to N\rho$, the available data are not
precise enough to discriminate between them. One can only put
constraint on their couplings versus the cut-off parameter
$\Lambda$, as shown by Figs.~2-4. If the data on both the two-pion
and radiative decays were improved significantly, one might
eventually hope to identify unambiguously the form of the $N^*$
coupling from a comparison of the two rates.

In conclusion, the $N^*(1535)N \rho$ vertex can be constrained by
the available experimental data from the radiative and two-pion
decays of $N^*(1535)$ resonance. The pure $S$-wave coupling gives
a good simultaneous fit to the data, though the large error bars
means that one cannot exclude either the pure vector or tensor
forms. The values of the coupling constant are strong in the sense
that they would predict a large $\rho$-exchange contribution to
$\eta$ production in nucleon-nucleon
scattering~\cite{geda,san,faldt,naka02,vetter} so that it would be
very unwise to neglect it.

%
%
\begin{acknowledgments}
The authors would like to thank Hai-qing Zhou and Wei Wang for
useful discussions. This work was partly supported by the National
Natural Science Foundation of China under grants Nos. 10435080,
10521003, and by the Chinese Academy of Sciences under project
No.~KJCX3-SYW-N2. One of the authors (CW) would like to thank the
Institute of High Energy Physics, CAS, for support and hospitality
during the initiation of this work.
\end{acknowledgments}
%
%

\end{document}